\documentclass[prl,twocolumn,aps,floatfix,longbibliography,natbib,nofootinbib,superscriptaddress]{revtex4-1}

\usepackage{pslatex,graphicx,dcolumn,bm,natbib,amssymb,amsmath,color,mathtools,mathrsfs,eufrak}

\usepackage[utf8]{inputenc}
\usepackage[T1]{fontenc}
\usepackage{hyperref}
\usepackage[english]{babel}
\usepackage{blindtext}
\usepackage{lipsum} 
\newcommand{\beq}{\begin{equation}}
\newcommand{\eeq}{\end{equation}}

\newcommand{\beqa}{\begin{eqnarray}}
\newcommand{\eeqa}{\end{eqnarray}}

\newcommand{\yw}[1]{\textcolor{black}{#1}} 

\usepackage{xr}
\makeatletter

\newcommand*{\addFileDependency}[1]{
\typeout{(#1)}
\@addtofilelist{#1}
\IfFileExists{#1}{}{\typeout{No file #1.}}
}\makeatother

\begin{document}


\title{Cell Division and Motility Enable Hexatic Order in Biological Tissues}

\author{Yiwen Tang}
\affiliation{Department of Physics, Northeastern University, Boston, MA 02115}
\author{Siyuan Chen}
\affiliation{Department of Physics, University of California, Santa Barbara, Santa Barbara, CA 93106}

\author{Mark J. Bowick}%
\affiliation{Department of Physics, University of California, Santa Barbara, Santa Barbara, CA 93106}
\affiliation{Kavli Institute of Theoretical Physics, University of California, Santa Barbara, Santa Barbara, CA 93106}

\author{Dapeng Bi}
\affiliation{Department of Physics, Northeastern University, Boston, MA 02115}

\begin{abstract}
Biological tissues transform between solid-like and liquid-like states in many fundamental physiological events. Recent experimental observations further suggest that in two-dimensional epithelial tissues these solid-liquid transformations can happen via intermediate states akin to the intermediate hexatic phases observed in equilibrium two-dimensional melting. The hexatic phase is characterized by quasi-long-range (power-law) orientational order but no translational order, thus endowing some structure to an otherwise structureless fluid.  While it has been shown that hexatic order in tissue models can be induced by motility and thermal fluctuations, the role of cell division and apoptosis (birth and death) has remained poorly understood,    despite its fundamental biological role. Here we study the effect of cell division and apoptosis on global hexatic order within the framework of the self-propelled Voronoi model of tissue.  Although cell division naively destroys order and active motility facilitates deformations, we show that their combined action drives a liquid-hexatic-liquid transformation as the motility increases. The hexatic phase is accessed by the delicate balance of dislocation defect generation from cell division and the active binding of disclination-antidisclination pairs from motility. We formulate a meanfield model to elucidate this competition between cell division and motility and the consequent development of hexatic order. 
\end{abstract}
\maketitle 

Organ surfaces are often covered with 2D confluent monolayers of epithelial or endothelial cells\yw{,} which provide functional separation from the surrounding environment. During development these cells grow, divide and move, dynamically reorganizing the entire tissue. Regulated by a complex set of chemical and mechanical signaling pathways\cite{lecuit2011,janmey2011,mammoto2010,tenney2009}, tissue frequently undergoes a transition from a structureless fluid-like state to a state capable of supporting a variety of stresses, most notably elastic stresses\cite{atiaGeometricConstraintsEpithelial2018,park2015,mongera2018,malinverno2017a,angelini2011,garcia2015,puliafito2012}. Such transformations have recently been analyzed as a crossover from a liquid to an amorphous solid\cite{bi2015,biMotilityDrivenGlassJamming2016}. 
In two-dimensional (2D) systems in equilibrium,  however, liquids can develop rigidity via two consecutive transitions, \yw{the first corresponding to the development of orientational order without translational order and the second adding translational order to the existing orientational order}\cite{halperin1978,nelson1979}.  The intermediate phase with (quasi-long-range) orientational order but translational disorder is known as the hexatic phase and has been shown to occur in a very wide variety of physical systems\cite{chou1998,murray1992,zahn1999,zahn2000,han2008,seshadri1991,knobler1992,huang1992,huang1993,podgornik1996,vanwinkle1997,rill1991,bishop1994,murray1990}. The hexatic is a particular type of structured fluid since it flows like a fluid but has orientational rigidity.  
 
Previous theoretical and computational models of dense tissues have studied the emergence of hexatic order, with focus on the effects of thermal fluctuations\cite{li2018,durand2019,guo_li_ai_phasefield} and 
motility\cite{pasupalakHexaticPhaseModel2020,loewe2020,paoluzzi2021,Li_Ai_njp_2021}.
\yw{Modeling typically studies the inverse process of disordering by melting from the crystalline state.}
 Realistic tissues, however, are very rarely crystalline with a few exceptions\cite{cohen2020,Hilgenfeldt_2008}. Cell division and apoptosis almost always destroy the crystalline state\cite{a.matoz-fernandez2017}
 and yet there has been no direct observation of the hexatic phase in \textit{in vitro} biological tissues, including those undergoing a solid-
 \yw{liquid}
transition\cite{park2015,malinverno2017a,atiaGeometricConstraintsEpithelial2018,mitchelPrimaryAirwayEpithelial2020}. 

Recent \textit{in vivo} experiments on Drosophila embryos have uncovered hexatic order during development with cell division\cite{kanesaki2011, wu2021}, along with the associated increase of orientational correlations\cite{kaiser2018}. The mechanism behind the emergence of this orientational order has remained unclear.

Here we analyze whether biological systems can exhibit this rather subtle phase by studying numerically and analytically the self-propelled Voronoi (SPV) model of cellular tissue including cell division and death\cite{biMotilityDrivenGlassJamming2016}. We compare a variety of structural properties with and without division, including translational and orientational order parameters, order field correlation functions in space, order field susceptibility, and topological defect densities. 
We find that the interplay of division/apoptosis and cell motility does indeed give rise to a hexatic regime. In the absence of cell division, the model undergoes a crystal-hexatic and hexatic-liquid
transition. With both cell division and motility, the model is driven through distinct liquid-hexatic and hexatic-liquid transitions with a re-entrant state, or phase, diagram. While cell motility is typically thought to disorder, we show that the combined effect of cell division and cell motility allows access to the hexatic state. A key role in this process is played by topological defects, both disclinations and dislocations. 

\paragraph{Model}
We model a 2D cell layer using the Self-Propelled Voronoi (SPV)\cite{biMotilityDrivenGlassJamming2016} version of the vertex model\cite{farhadifar2007,li2019a,liBiologicalTissueinspiredTunable2018,yan2019,mitchelPrimaryAirwayEpithelial2020,das2021}. 
The cell shapes and the cellular network are determined based on the Voronoi tesselation\cite{dirichlet1850,honda1978} of the cell centers  $\{ \vec{r}_i \}$. Here mechanical interactions in the tissue are controlled by the energy functional $ E = \sum_{i=1}^N [K_A(A_i-A_0)^2+K_P(P_i-P_0)^2]$. 
The first term,  quadratic in the cell areas $\{A_i\}$,  originates from the incompressibility of cell volume, giving rise to a 2D  area elasticity constant $K_A$ and preferred area $A_0$\cite{farhadifar2007,staple2010a}. 
The second term quadratic in the cell perimeters $\{P_i\}$  arises from the contractility of the cell cortex, with an elastic constant $K_P$\cite{farhadifar2007}. Here $P_0$ is the target cell perimeter\cite{bi2015}, representing the interfacial tension set by the competition between the cortical tension and the adhesion between adjacent cells\cite{staple2010a}.
The target shape index $p_0 = P_0 /\sqrt{A_0}$ effectively characterizes the competition between cell-cell adhesion and cortical tension, acting as a signature for the solid-liquid phase transition\cite{bi2015,yan2019}.
Apart from the effective mechanical interaction force $\mathbf{F}_i=-\nabla_i E$, cells are self-propelled. A self-propulsion force is exerted along the cell polarity direction $\mathbf{\hat{n}_i}=(\cos\theta_i, \sin\theta_i)$, where $\theta_i$ is the polarity angle.  The self-propulsion has a  constant magnitude $v_0/\mu$, with the inverse of a frictional drag $\mu$. The  equation of motion for each cell is given by
\begin{equation}
    \dot{ \vec{r_i}}  = \mu \vec{F_i} + v_0\mathbf{\hat{n}_i}.
    \label{eq:eom}
\end{equation}
The polarity angle obeys rotational diffusion: $d \theta_i / dt =\eta_i(t) $, where $\eta_i(t)$ is white-noise ($\langle \eta_i(t)\eta_j(t')\rangle  =2 D_r \delta(t-t')\delta_{ij}$), with $D_r$ the rotational diffusion rate. 

In addition to the polarized self-propulsion, cell division and apoptosis serve as another source of active forcing in living tissues\cite{a.matoz-fernandez2017,doostmohammadi2015,cislo2023, saw2017}.  
\yw{In the SPV model, every cell has an equal division rate $\gamma_0$.}
For each cell division, a daughter cell is introduced by randomly seeding a point at a distance of  $d=0.1$ (in units of the average cell diameter) near the mother cell. 
%
In order to study the density-independent effects of cell-division, we keep the number density of the tissue constant by implementing apoptosis at the same rate as division. Apoptosis is then performed on randomly chosen cells, which removes the cells from the tissue. This simulation scheme mimics the maintenance of homeostatic balance in a tissue\cite{basan2011,ranft2010}.

The model can be nondimensionalized by expressing all lengths in units of $\sqrt{\bar{A}}$, where $\bar{A}$ is the average cell area in the tissue and time in units of $1/(\mu K_A\bar{A})$. Three independent parameters remain the cell division/apoptosis rate $\gamma_0$, the magnitude of motility $v_0$, and the cell shape index $p_0$. Throughout the simulations, we choose $D_r=1$ without loss of generality. 

The confluent tissue with $N$ cells is simulated in a square box with size $L =\sqrt{N}$ under periodic boundary conditions. We numerically simulate the model using the open-source software cellGPU\cite{sussman2017}. The simulations start with a crystalline initial state in which cell centers form a triangular lattice. 
Eq.~\ref{eq:eom} is numerically integrated for $2\times10^6$ steps at a step size of $\Delta t = 0.05$. 
For all data presented, the analysis is based on the steady-state regime of the simulations (final $5\times10^5$ steps). In the supplementary material (Fig.\ref{fig:psi6_v0_ran_cry}
), 
\yw{we also simulate the model starting from amorphous states and do the cooling experiment to demonstrate that the results are independent of the initial condition and simulation approach.}
We set $p_0=3.6$ in our simulations. During the melting process, the tissues undergo a transition from crystalline to hexatic to liquid as motility increases\cite{pasupalakHexaticPhaseModel2020}.

\paragraph{Signature for the emergence of hexatic order}
 Translational and orientational symmetries distinguish the three phases crystalline, hexatic and liquid. A 2D crystalline phase has quasi-long-range translational order and long-range orientational order while the liquid phase has no long-range order of either kind. These two symmetries are related but not concomitant. The system in the hexatic phase has no long-range translational order but retains quasi-long-range orientational order \cite{halperin1978, nelson1979}.   

\begin{figure}
    \centering
    \includegraphics[width=\linewidth]{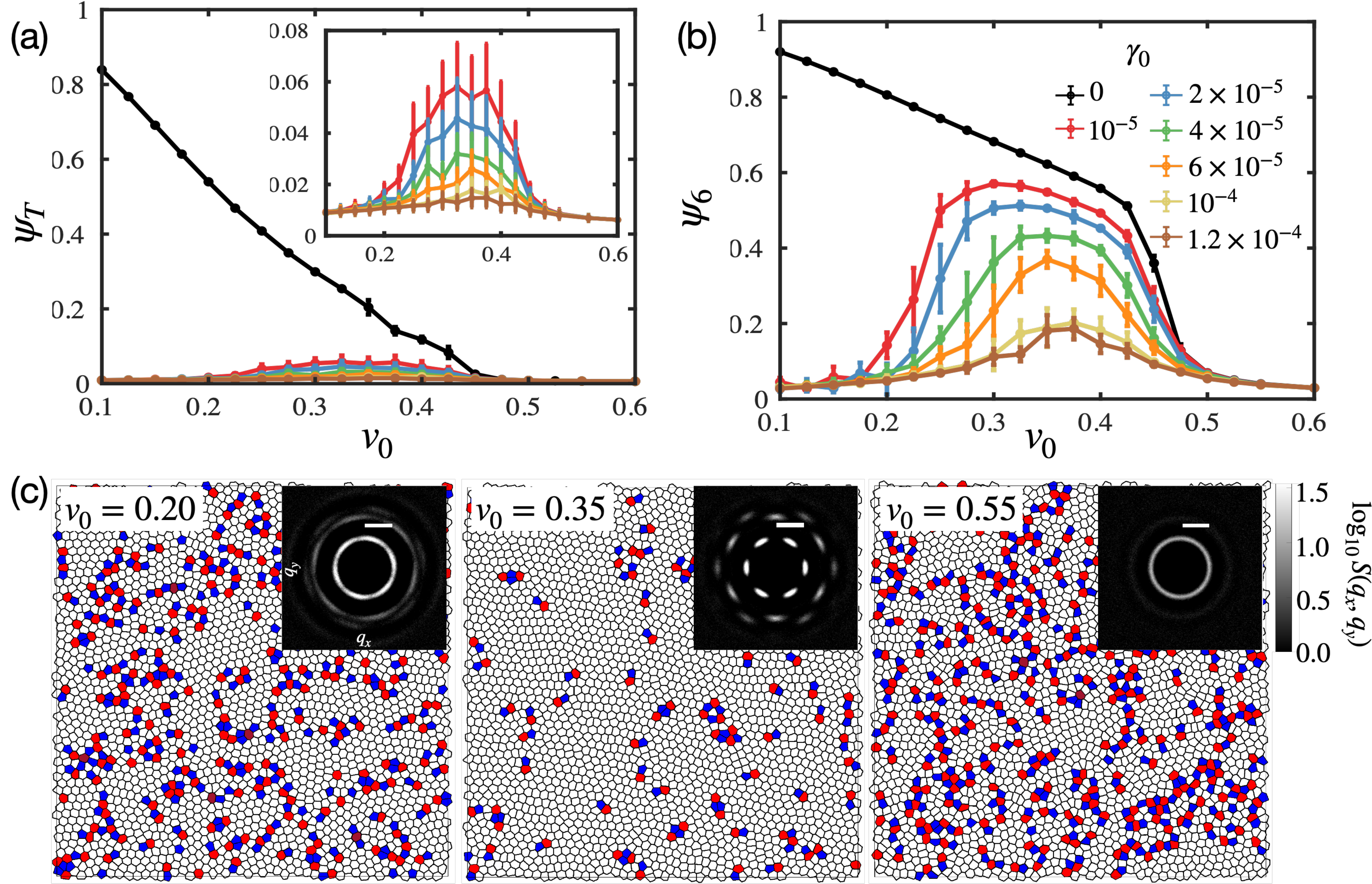}
    \caption{(a)  The translational order parameter  $\Psi_T$  and (b) the orientational order parameter  $\Psi_6$ as a function of the cell motility $v_0$ at various division rates $\gamma_0$. \yw{The errorbar represents the standard deviation over the ensemble of random simulations.}
    (c) \yw{The snapshot and structure factors $S(\bf{q})$ plotted for various $v_0$ at fixed division rate $\gamma_0=2\times10^{-5}$. Blue represents cells with disclination charge $q_i=1$, red represents cells with $q_i=-1$, and dark red represents cells with $q_i=-2$. The $v_0=0.35$ snapshot has dislocation clusters but no disclinations.}
    }
\label{fig:psi_v0}
\end{figure}

 We begin by quantifying translational and orientational order at the tissue level.  
 The translational order is quantified by $\psi_T(\vec{r}_j)=\exp{(i\vec{G}_r\vec{r}_j)}$, where $\vec{G}_r$ represents a reciprocal vector in reciprocal space.
The orientational order is quantified 
$\psi_6(\vec{r}_j)=\left( 1/\sum_{i=1}^{z_j}l_{ij} \right) \sum_{i=1}^{z_j}l_{ij}\exp{(i6\theta_i^j)}$,
where the sum runs over the $n$ Voronoi neighbors of the cell and is weighted by their shared edge length\cite{mickel2013,armengol-collado2022d,armengol-collado2022e}. $\theta_i^j$ is the angle of the neighboring joint vector $(\vec{r}_i-\vec{r}_j)$ to a reference axis. 
In Fig.~\ref{fig:psi_v0}, we plot the tissue-level order parameters $\Psi_6= \frac{1}{N}\sum_{j=1}^N \psi_6(\vec{r}_j)$ and  $\Psi_T= \frac{1}{N}\sum_{j=1}^N \psi_T(\vec{r}_j)$ as a function of $v_0$.
In the absence of cell division (black lines), the tissue is a crystal at low $v_0$ where both $\Psi_T, \Psi_6$ are close to 1.  The order parameters decrease monotonically with increasing $v_0$. For $0.35 \lesssim v_0 \lesssim 0.45 $,  the tissue lacks translational order but retains orientational order, suggesting the existence of a hexatic phase before melting into a liquid phase at higher $v_0$. This result is consistent with the \emph{crystal-hexatic-liquid} melting scenario in the previous study using a similar model\cite{pasupalakHexaticPhaseModel2020}.

When cells divide (color lines in Fig.~\ref{fig:psi_v0}), $\Psi_T$ is always close to zero at any value of $v_0$. This clearly illustrates that activity due to cell cycling (division/death) always destroys the translational order and therefore forbids the formation of permanently frozen structures\cite{a.matoz-fernandez2017}.   Remarkably, while an actively dividing tissue lacks translational order, it retains orientational order for a large range of $v_0$ values. This suggests the emergence of a hexatic phase at intermediate $v_0$ values. A transition from liquid to hexatic to liquid is visualized by the structure factor $S(\bf{q})$ for various $v_0$ at a fixed division rate.
 
\begin{figure}
    \centering
    \includegraphics[width=\linewidth]{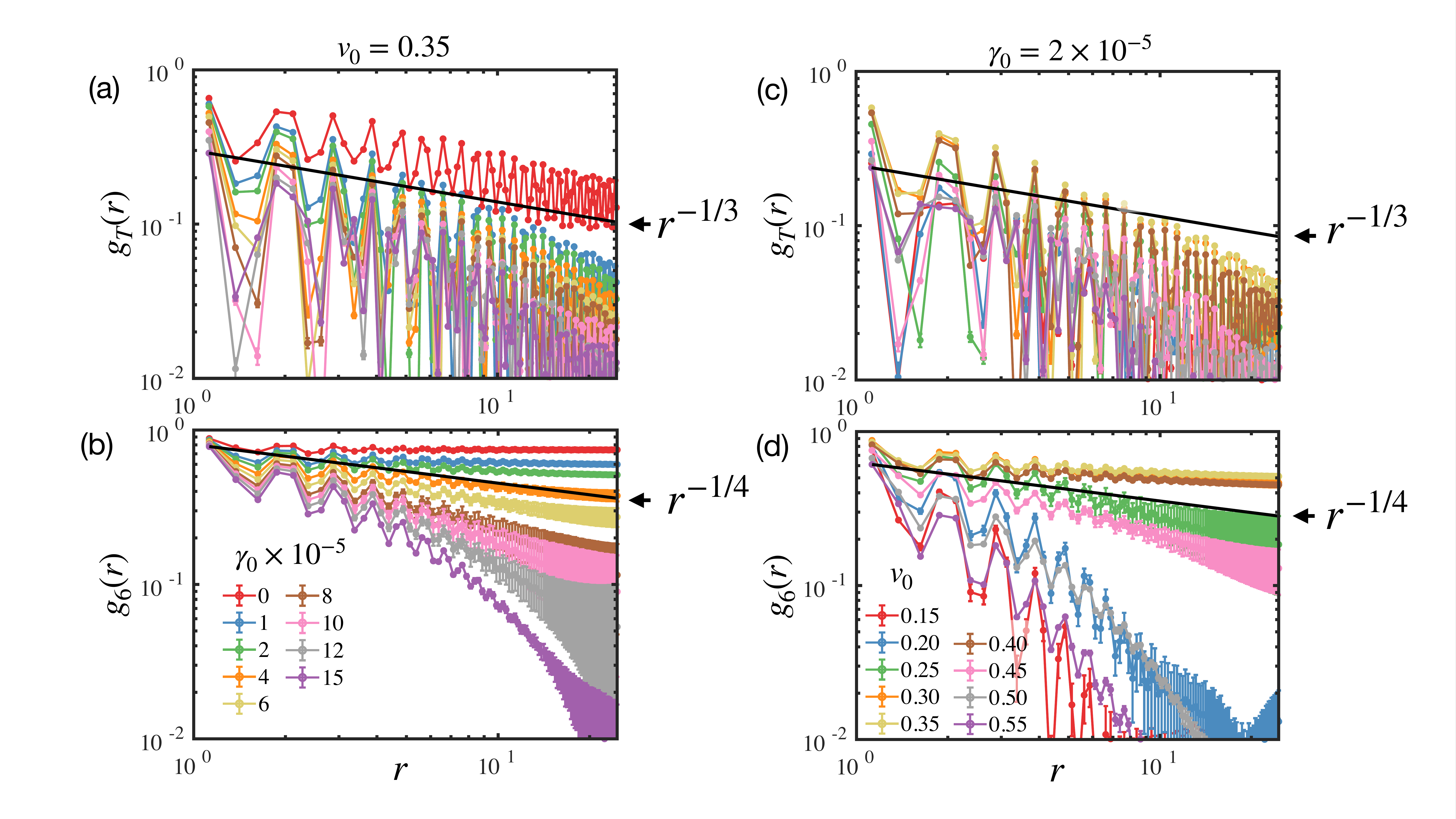}    
    \caption{The (a) translational and (b) bond-orientational correlation functions at intermediate cell motility $v_0=0.35$. The (c) translational and (d) bond-orientational correlation functions at low cell division rate $\gamma_0=2\times10^{-5}$.}
\label{fig:gr_div}
\end{figure}

In order to determine the location of the transitions between different phases, we next compute the bond-orientational and translational correlation functions. They are given by $g_\alpha(r)= \langle\psi_\alpha^*(r)\psi_\alpha(0)\rangle$,
with $r=|\vec{r}_i-\vec{r}_j|$ and $\alpha = 6,T$ corresponding to orientational order and translational order, respectively. 
The peaks of correlations are fitted by a power law decay $g_\alpha(r)\sim r^{-\eta_\alpha}$ (long-range order) and an exponential decay $g_\alpha(r)\sim e^{-r/\xi_6}$ (short-range order). KTHNY theory\cite{halperin1978,nelson1979,young1979,kosterlitz1973,bernard2011}  predicts $\eta_6=1/4$ at the hexatic-liquid transition point and $\eta_T=1/3$ at the crystal-hexatic transition point\cite{nelson2002, deutschlander2014}. 

The correlations are drawn and compared with reference power-laws ($\eta_T=1/3$ or $\eta_6=1/4$) in Fig.\ref{fig:gr_div} and  
Fig.\ref{fig:gr_no_div}. Melting (without cell division) allows quasi-long-range translational order at low $v_0$, decaying as a power law with $\eta_T \leq 1/3$. The translational order with cell division decays faster. Cell division also promotes the decay of bond-orientational correlations, but the low $\gamma_0$ still allows for quasi-long-range $g_6(r)$ with $\eta_6\leq1/4$ at intermediate $v_0$ values. A broken translational symmetry without broken orientational symmetry characterizes the emergence of a hexatic state. Exponential decay fits the orientational order better in both low- and high-$v_0$ liquid phases. 

\begin{figure}
    \centering
    \includegraphics[width=1\columnwidth]{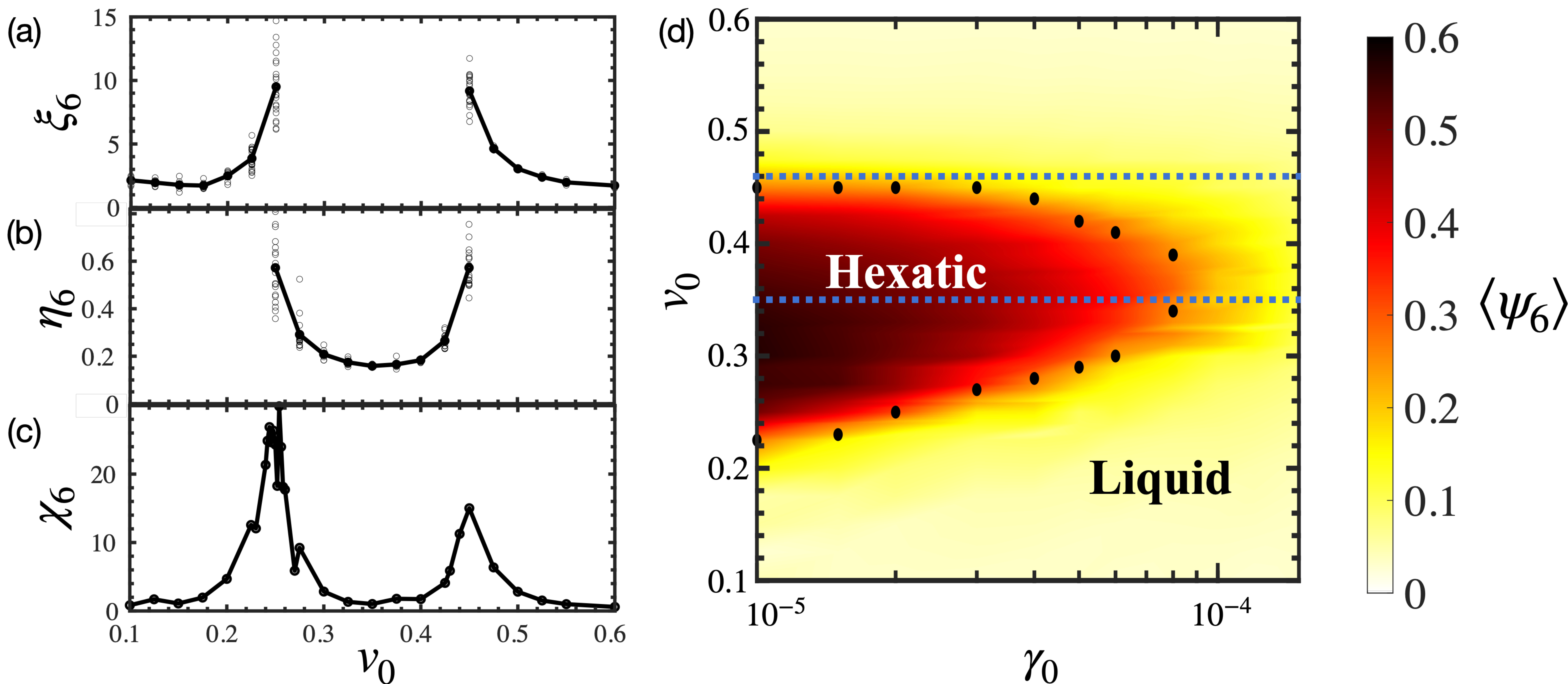}    
    \caption{
    (a) The correlation length $\xi_6$ and 
    (b) the power-law decay exponent  $\eta_6$ of the orientational correlation function are shown as functions of $v_0$ at constant $\gamma_0=2\times 10^{-5}$. 
    Circles represent the fitting exponents for different seeds, and the solid lines average the seeds. 
    (c) The hexatic order parameter susceptibility $\chi_6$. 
    (d) Phase diagram as a function of cell division rate $\gamma_0$ and motility $v_0$.
    } 
    \label{fig:xi_eta_chi6}
\end{figure}

The fitted exponents $\eta_6$ and $\xi_6$ at fixed division rate $\gamma_0=2\times10^{-5}$ are shown in Fig.\ref{fig:xi_eta_chi6} and in Fig.~\ref{fig:xi_eta_nodiv} for the case of no cell division.
These results confirm the emergence of two distinct liquid-hexatic and hexatic-liquid transitions when there is cell division. The correlations in the hexatic indeed display quasi-long-range order, well-fitted by power-law decays,  $g_6(r)\sim r^{-\eta_6}$, while outside the hexatic region correlations are short-range and well-fitted by exponential decays $g_6(r)\sim e^{-r/\xi_6}$.
As the hexatic phase is approached from either side, $\xi_6$ grows rapidly, consistent with a diverging correlation length.

Despite excellent agreement with the KTHNY model, the correlation functions and the associated quantities ($\xi_6, \eta_6$) near the onset of hexatic states suffer from large sample-to-sample variations as shown in Fig.~\ref{fig:xi_eta_chi6}(a,b). We have confirmed that this is not due to finite-size effects since, even at large system sizes, the behavior of $g_6(r)$ can range from exponential decay to a power-law decay (Fig.~\ref{fig:g6r_Ns_seeds}). Consequently, ($\xi_6$, $\eta_6$) cannot be used to pinpoint the precise location of the liquid-hexatic and hexatic-liquid transitions. 

We next take advantage of the large fluctuations that arise near critical points by using the order parameter susceptibility to pinpoint the transitions. 
The susceptibility is given by $\chi_\alpha=N(\langle|\Psi_\alpha|^2\rangle-\langle|\Psi_\alpha|\rangle^2)$, which
characterizes the fluctuations in the  translational ($\chi_T$) and orientational ($\chi_6$) order parameters. 
Since $\chi_\alpha$ is essentially an integral of the correlation function, it is expected to be more robust to finite-size or finite-time effects\cite{han2008,durand2019}. 

In the melting process without cell division (shown in Fig.\ref{fig:chi_no_div}), there is a sharp divergence of $\chi_T$ 
at $v_0=0.35$, indicating a crystal-hexatic transition. On the other hand, $\chi_6$ diverges at  $v_0 = 0.46\pm0.01$, which corresponds to the hexatic-liquid transition. By analyzing system sizes from $N=2430$ to $38880$, we show that these divergences are robust to finite-size effects (Supplemental Material).

In contrast, $\chi_6$  with cell division (at  $\gamma_0=2\times 10^{-5}/s$) generate two peaks ( Fig.\ref{fig:xi_eta_chi6}(c)). The divergence of $\chi_6$ determines two distinct transition points at $v_0=0.25 \pm 0.01$ and at $v_0=0.45 \pm 0.01$. Whereas the second point is a vestige of the hexatic-liquid transition in the absence of cell division, the first transition point arises solely from cell division.  
Here, a state that would be crystalline in the absence of cell division becomes hexatic when cells divide.  

Exploring various cell division rates $\gamma_0$ and active motilities $v_0$ at fixed $p_0=3.6$, we plot the $v_0-\gamma_0$ phase diagram in Fig.\ref{fig:xi_eta_chi6}(d). 
Color indicates the mean magnitude of the global orientational order over tens of thousands of frames. 
Black dots mark the peaks of $\chi_6$ at various division rates.
The two transition points approach each other and disappear as the division rate increases. \yw{The blue dashed lines indicate the liquid-hexatic and hexatic-crystal transition points in the absence of cell division.}
\yw{We also investigate the $p_0$ dependence of the phase diagram in Fig.\ref{fig:psi_v0_p0} and compare it to previous results by Pasupalak et al.\cite{pasupalakHexaticPhaseModel2020} which do not have cell division.}

\paragraph{Disclinations and Dislocations}

\begin{figure}
    \centering
    \includegraphics[width=1\columnwidth]{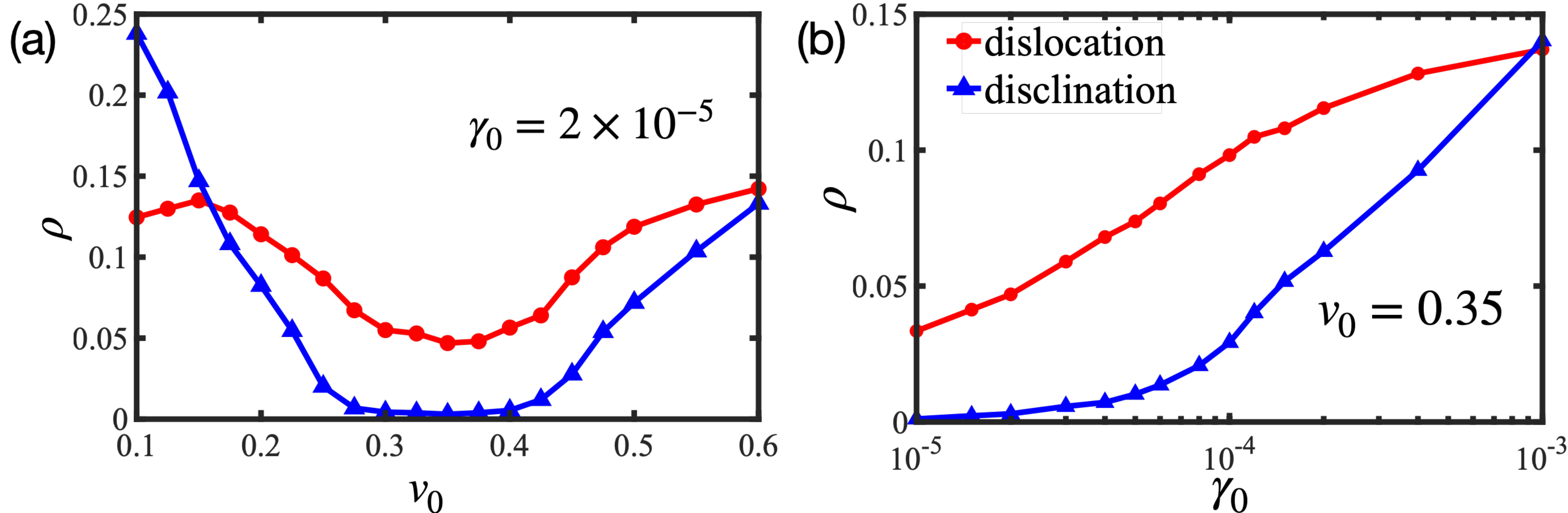}
    \caption{
    (a) The volume densities of dislocations and disclinations are plotted as functions of $v_0$ at constant cell division rate  $\gamma_0=2\times10^{-5}$.  (b) The same quantities are plotted at a constant  $v_0=0.35$ and varying $\gamma_0$.
    }
    \label{fig:defect_div}
\end{figure}

According to KTHNY theory\cite{halperin1978,nelson1979,young1979,kosterlitz1973,bernard2011}, the distinct phases crystalline, hexatic and liquid are characterized by the distributions of the basic topological defects known as disclinations and dislocations. Whereas the pure crystalline phase is defect free, or equivalently all defects are tightly bound in defect-antidefect pairs, the hexatic phase has a non-vanishing density of free dislocations and the liquid phase has a non-vanishing density of free disclinations.

We can define a  charge $q_i=6-z_i$\cite{bowick2009} associated with disclinations, where  $z_i$ is the coordination number (number of neighbors) of the $i$th cell. Hexagonal cells are thus "neutral", pentagonal cells have charge +1, heptagonal cells charge -1 and so on.  Dislocations, the defects that disrupt translational order but preserve orientational order, correspond to tightly bound $5-7$ pairs. They are neutral as disclinations but possess a net vectorial charge, the Burgers vector. We approximate the Burgers vector by the displacement vector separating the 5 and the connected 7. In general there will be clusters of connected defects and one must measure the associated disclination and dislocation charges of the entire cluster. The density of disclinations and dislocations are calculated by their volume fraction averaged over time. 
 
As shown in Fig.\ref{fig:defect_div}(a), cell division creates dislocations at a rate dependent on motility. Division tends to disorder, favoring a liquid. What about motility? At low motility, division disordering wins. At high motility both processes \yw{generate} disorder, leading again to a liquid. But for a significant range of intermediate motilities, we see that the number density of free disclinations falls to zero whereas the free dislocation density is finite. How is this possible? In this intermediate region we hypothesize that disclinations are able to explore sufficient configuration space to access local free energy minima at which all disclinations find their anti-disclinations and bind into dislocations, thus leading to a hexatic. Fig.\ref{fig:defect_div}(b)
 shows the defect density dependence on cell division rate at a fixed motility in the hexatic regime, showing that sufficiently high division rates lead to a non-zero density of free disclinations, thus melting the hexatic to a liquid.
\yw{The middle snapshot in Fig.\ref{fig:psi_v0}(c)} is a representative snapshot of a hexatic state ($v_0=0.35$, $\gamma_0 = 2\times 10^{-5}$). Note the presence of dislocation complexes but no isolated disclinations. Movie.~S1
shows a dynamic evolution of states with various values of cell motility at a fixed division rate. 
The densities of dislocations and disclinations are indicated by color as a function of cell division rate $\gamma_0$ and motility $v_0$ at fixed $p_0=3.6$ in Fig.~\ref{fig:defect_v0_gamma0}. Black dots mark the same data in Fig.~\ref{fig:xi_eta_chi6}(d).

\paragraph{Meanfield Model}

\begin{figure}    
    \centering
    \includegraphics[width=\linewidth]{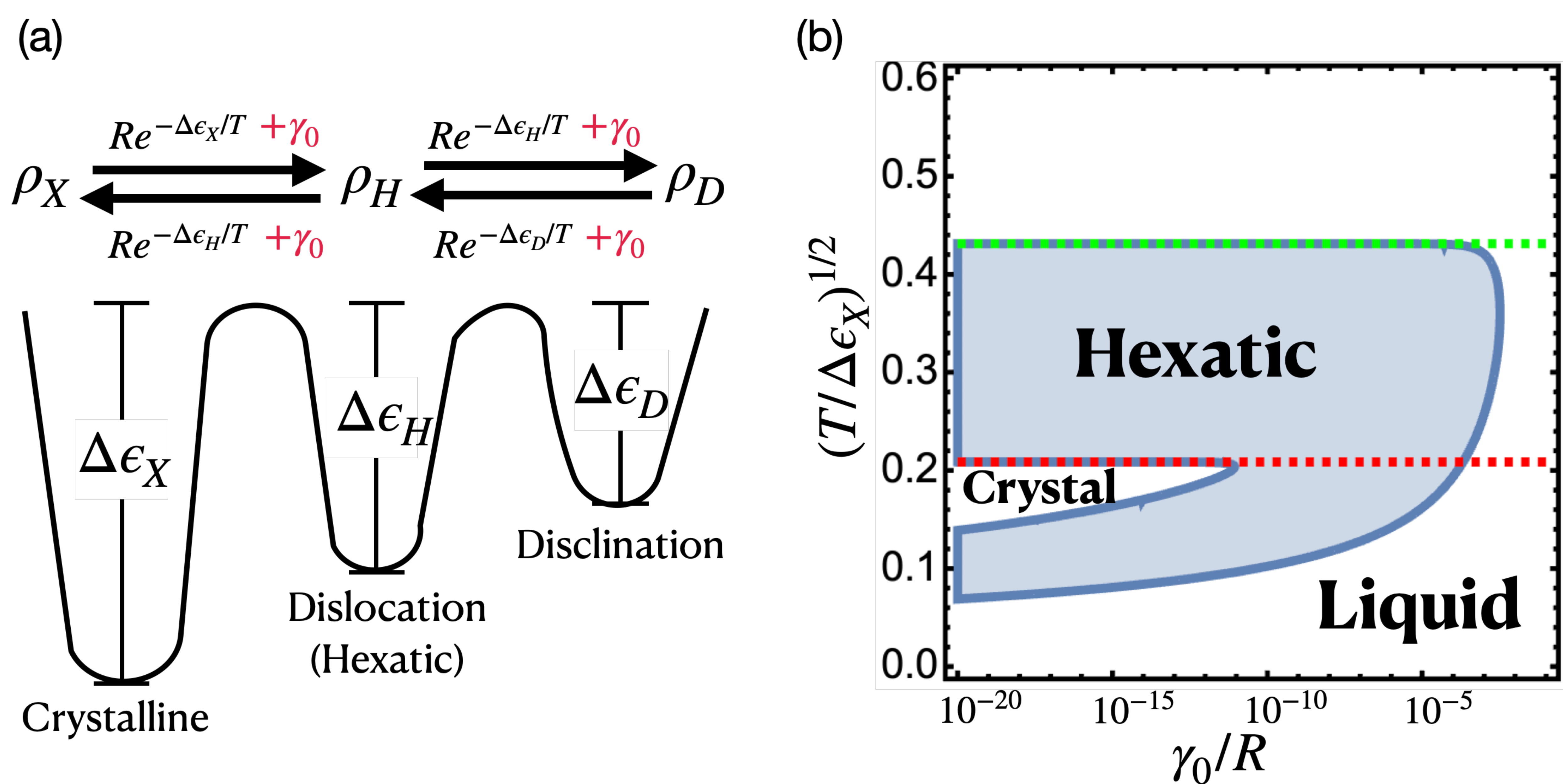}
    \caption{A meanfield description for defect dynamics in a tissue. (a) The energy landscape and transition diagram between states. (b) \yw{The phase diagram of meanfield model as the function of the dimensionless division rate $\gamma_0/R$ and motility $(T/\Delta\epsilon_X)^{1/2}$. The green(red) dashed line indicates the liquid-hexatic and hexatic-crystal transition in the absence of cell division.}
    }   
    \label{fig:mean_field}
\end{figure}

To further understand the emergence of hexatic order through cell division we develop a simple meanfield model (MF) (Fig.\ref{fig:mean_field}) incorporating the competition between cell division and motility.  We simplify the state of a small cell cluster ($\sim 4$ cells) using a meanfield approximation that allows three states: (a) crystalline \yw{solid} state ("ordered"), (b) an isolated single dislocation, and (c) an isolated single disinclination. Transitions between states arise from fluctuations over the energy barriers $\Delta\epsilon_i$, as illustrated in Fig.~\ref{fig:mean_field}(a). Fluctuations arise from both Brownian motility forces and cell division, leading, in the low temperature/velocity limit, to an equal density distribution of states rather than only the "ordered" state (see Supplementary Text).
In the steady-state limit\yw{(Fig.\ref{fig:mean_field}(b)), the asymmetric boundary of the hexatic region (determined by the threshold of $\rho_H$ and $\rho_D$) is remarkably similar to our simulations and the phase diagram shows a re-entrant liquid-hexatic-liquid transition with changing motility, as in Fig.~\ref{fig:xi_eta_chi6}(d).
The MF model predicts unique behavior for ultra-low division rates ($ \gamma_0/R \ll 10^{-10}$), where tissues undergo a complex temperature-dependent transition sequence, following a Liquid-Hexatic-Crystal-Hexatic-Liquid path at constant $\gamma_0/R$ (Fig.\ref{fig:mean_field}(b) and also see Supplementary Text for a detailed discussion). Remarkably, our phase diagram closely mirrors that of 2D melting on a random substrate~\cite{sachdev1984}. In both models, temperature drives phase transitions, and in our case, cell division plays a role analogous to substrate disorder, introducing persistent, random spatial distortions.}

\paragraph{Discussion}
 
The subtle balance required to establish hexatic order in equilibrium means that it is often confined to a  rather narrow region of the relevant parameter space. Our findings suggest that cell division provides a new way of exploring the configuration space of physical systems, as noted above. In particular, the dynamics of dislocation defects generated by cell-division, both self-propelled and relaxational, promote fluctuations over barriers separating the hexatic phase from crystalline or liquid phases. This phenomenon, which we may call defect-driven structure development, may well have implications beyond biological systems. In terms of the configuration space explored by the vertex model, cell division and apoptosis correspond to adding T2 moves (or interstitial insertion/deletion) to the allowed lattice updates -- this yields a more efficient exploration of the space of all Voronoi tesselations and thus better routes to local hexatic minima \cite{bowick2007,irvine2012,j.bowick2007}. It is remarkable that the early work of Swope and Andersen\cite{swope1995d} found the hexatic phase by employing a grand canonical ensemble in which particles are added and removed.  The mechanism we find here is very different from that found in colloids\cite{han2008} and models of active particles\cite{digregorio2018}, where packing density plays a crucial role.

We have taken cell division to be isotropic. The inclusion of oriented cell divisions, however, would only enhance hexatic order. Recent work\cite{cislo2023} has shown that oriented cell divisions can give rise to novel four-fold orientational order {\it in vivo} through active defect climb, where defects introduced into the nascent lattice by cell divisions are healed by subsequent divisions along a well-defined global polarity axis. The effect of oriented divisions on hexatic order is a subject for the future.

\begin{acknowledgments}  
This work was supported in part by NSF DMR-2046683 (Y.T. and D. B.),  PHY-1748958 (D. B. and M. J. B.), the Center for Theoretical Biological Physics NSF PHY-2019745 (Y. T. and D. B.), Alfred P. Sloan Foundation (Y. T. and D. B.) and The Human Frontier Science Program (Y. T. and D. B.))
\end{acknowledgments}  

\bibliography{MyLibrary}

\appendix

\section{Supplementary Text}
\section{Susceptibility of Order Parameters}$\chi_6$ is a measure of the fluctuation in tens of thousands of frames, which come from 50 simulations. Each simulation has a different seed, controlling the initial velocity direction and dividing cells. $\chi_T$ is a measure of the fluctuations of the time-averaged $\Psi_T$ from the 20 simulations. The ensemble average plays a key role in the susceptibility calculation.

\section{ Tracking Topological Defects}
The video (Movie.~S1\label{mov:defects}) shows a dynamic evolution of states with various cell motilities ($v_0=0.15,0.35,0.55$) with a fixed division rate ($\gamma_0=2\times 10^{-5}$). We color the dislocations in cyan($z_i<6$) and magenta($z_i>6$) and color the disclinations in blue($z_i<6$) and red($z_i>6$).
At $v_0=0.15$(left), the fluctuations from motility are insufficient to anneal the dislocations created by cell division. Dislocations eventually unbind into disclinations. The tissue is in a liquid state. $v_0=0.35$(middle) allows dislocations to reorganize and overcome barriers, leading to a hexatic. In contrast $v_0=0.55$(right) generates the dislocations and disclinations by itself.


\begin{figure}[htbp]
    \centering
    \includegraphics[width=0.5\textwidth]{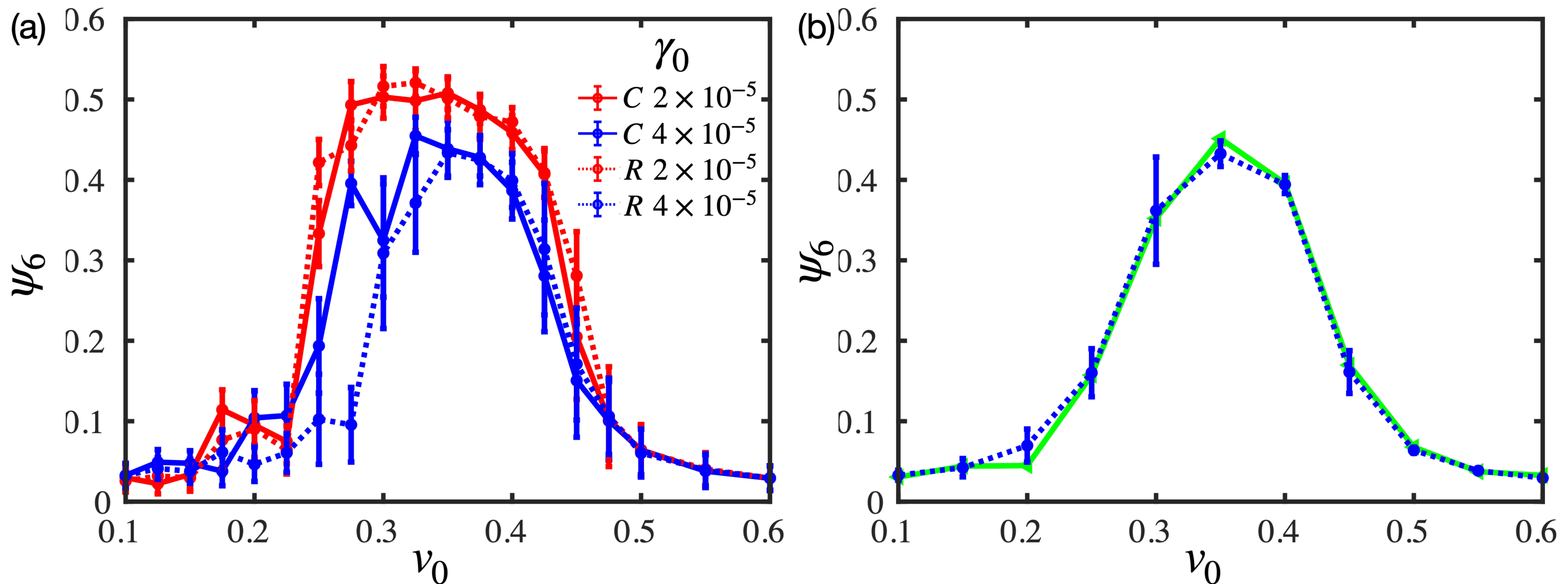}
    \caption{\yw{\textbf{The orientational order parameter is independent of simulation protocols.}
    (a) The orientational order parameter remains consistent across different initial conditions. The crystal (solid lines) and random (dash lines) initial conditions both lead to the same form for $\Psi_6(v_0)$. $\Psi_6$ is non-zero at intermediate $v_0$ values and declines for both higher and lower $v_0$. 
    (b) The orientational order parameter is unaffected by the approach taken during the simulation process. The cooling run (green solid line) has the same $\Psi_6(v_0)$ function as the direct run (blue dash lines) from crystal initial conditions at constant division rate $\gamma_0=4\times10^{-5}$. In the cooling run, we start from the simulation for $v_0=0.60$. Subsequently, we use the final state of $v_0=0.6$ as the initial state for $v_0=0.55$ and employ the resulting state of $v_0=0.55$ as the starting state for $v_0=0.50$, continuing this sequence until we reach $v_0=0.10$.}} 
    \label{fig:psi6_v0_ran_cry}
\end{figure}

\begin{figure}[htbp]
    \centering
    \includegraphics[width=0.4\textwidth]{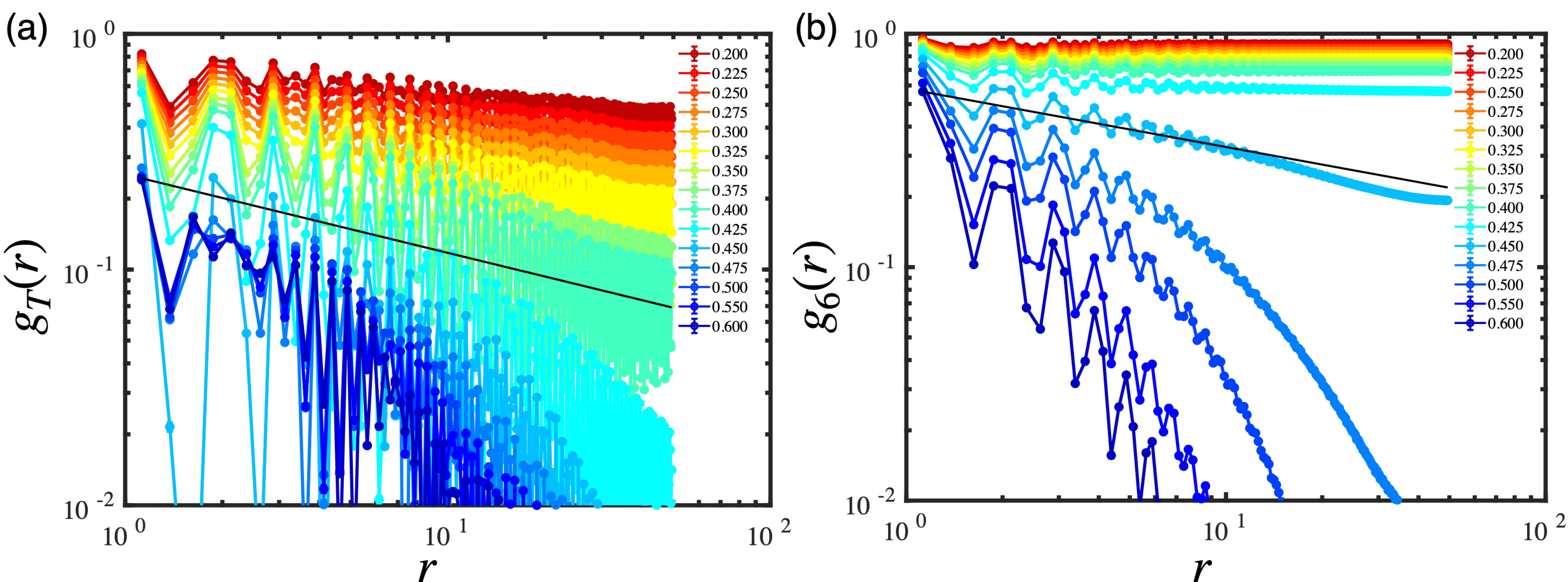}
    \caption{\textbf{The (a) translational and (b) bond-orientational correlation functions in the absence of cell division.}
    The curves evolve from a power-law decay to an exponential decay as $v_0$ increases. (a) The reference line $g_T \sim r^{-1/3}$ indicates the theoretically expected for the crystal-hexatic transition, while (b) the $g_6 \sim r^{-1/4}$ fall-off indicates the expected decay for a hexatic-liquid transition. The decay of $g_6$ lags $g_T$, leaving an intermediate range of $v_0$ without quasi-long-range translational order but with quasi-long-range orientational order.}
    \label{fig:gr_no_div}
\end{figure}

\begin{figure}[htbp]
    \centering
    \includegraphics[width=.4\textwidth]{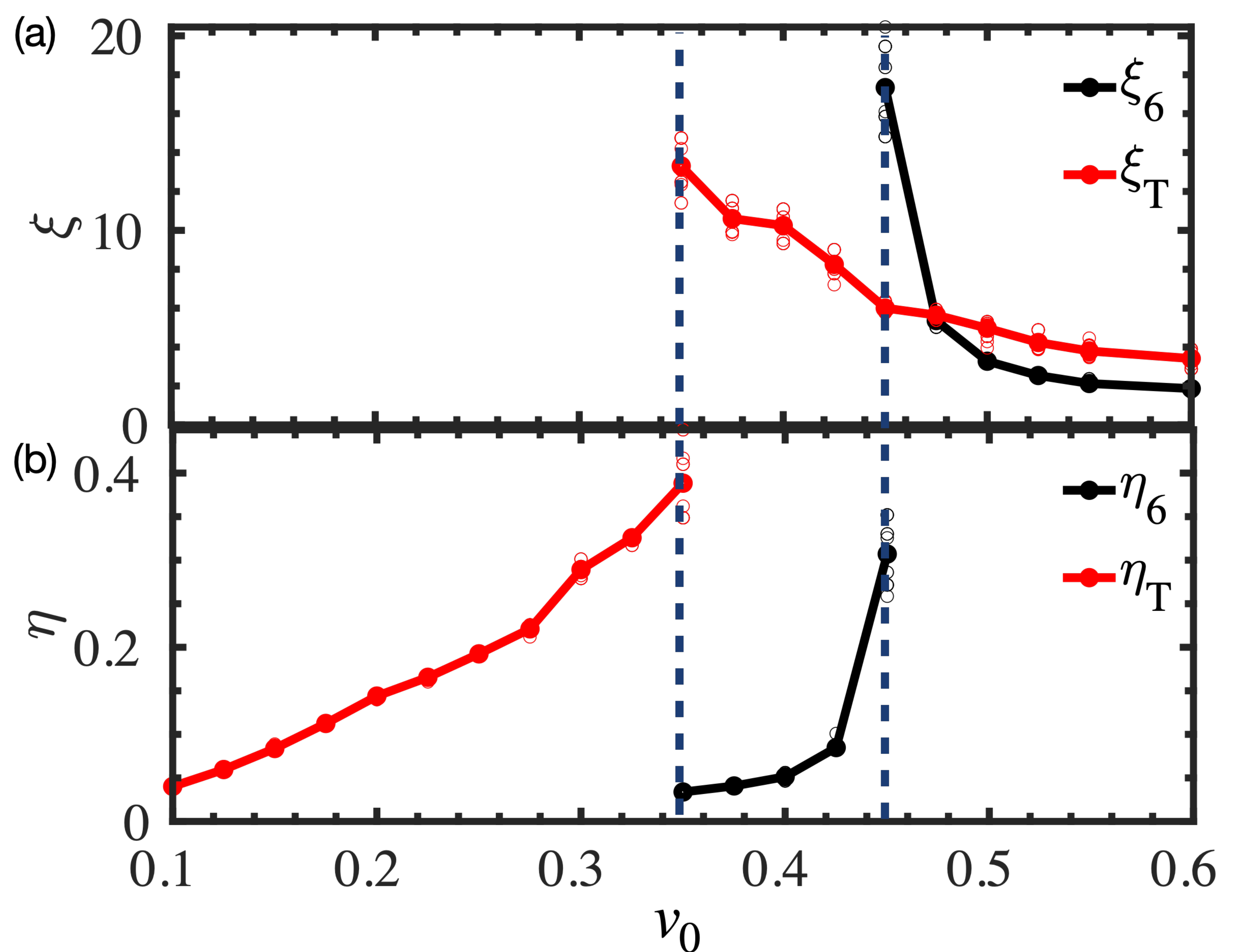}
    \caption{\textbf{The correlation length $\xi_\alpha$ and exponent $\eta_\alpha$ ($\alpha=6,T$) as a function of $v_0$ in the absence of cell division.}  The translational correlations in the crystal phase are quasi-long-range with a power-law decay $g_T(r)\sim r^{-\eta_T}$,  and those in the hexatic and liquid range are short-range with an exponential decay $g_T(r)\sim e^{-r/\xi_T}$. The orientational correlations in the hexatic are quasi-long-range with a power-law decay $g_6(r)\sim r^{-\eta_6}$, and those in the liquid are short-range with an exponential decay $g_6(r)\sim e^{-r/\xi_6}$. 
    $\xi_6$ and $\eta_6$ grow significantly at the transition point. 
    }
    \label{fig:xi_eta_nodiv}
\end{figure}

\begin{figure}[htbp]
    \centering
    \includegraphics[width=0.5\textwidth]{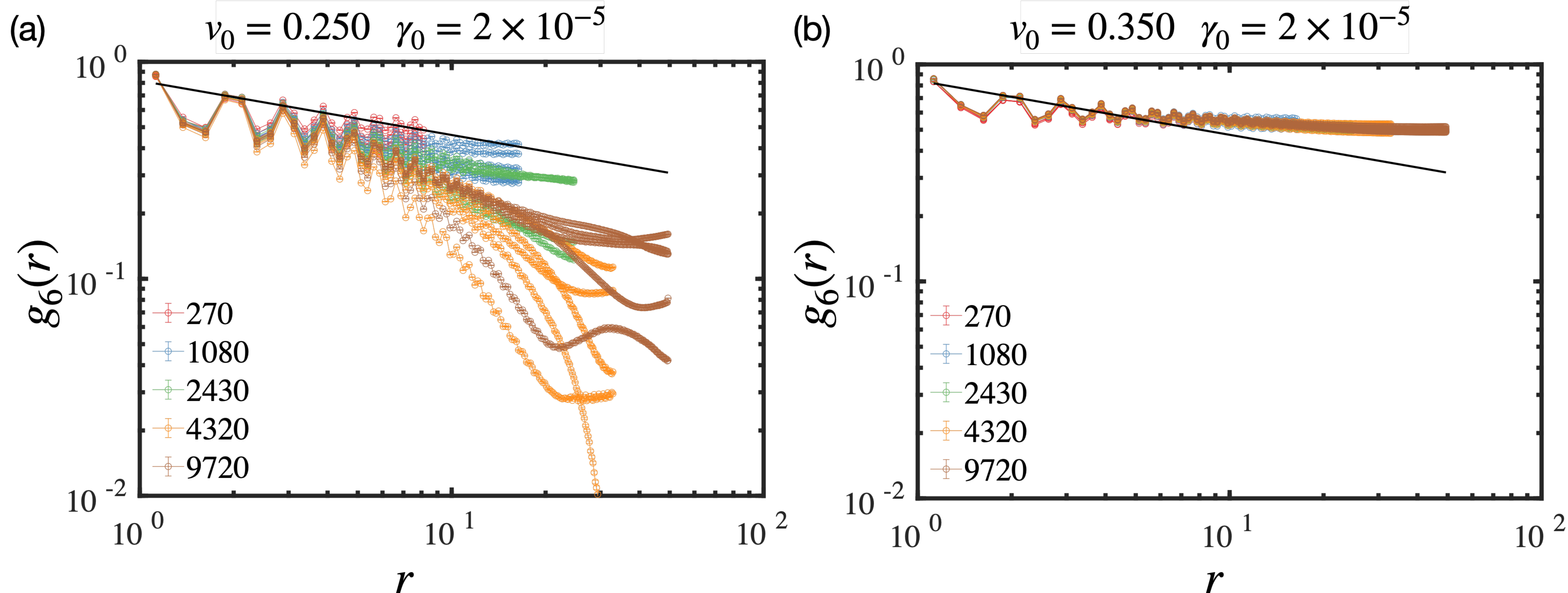}
    \caption{\textbf{The finite-size effects for the translational and orientational correlations.} Different colors represent various $N$ and different lines indicate simulations with distinct seeds. (a) Near the transition point ($v_0=0.25$ and $\gamma_0=2\times10^{-5}$), the behavior of $g_6(r)$ suffers large sample-to-sample fluctuations, ranging from exponential decay to a power-law decay.
    (b) Deep in the hexatic phase ($v_0=0.35$ and $\gamma_0=2\times10^{-5}$), the finite-size analysis over a large range of $N$ confirms quasi-long-range order. The power-law decay is independent of system size and seeds.
    }
    \label{fig:g6r_Ns_seeds}
\end{figure}

\begin{figure}
    \centering
    \includegraphics[width=0.4\textwidth]{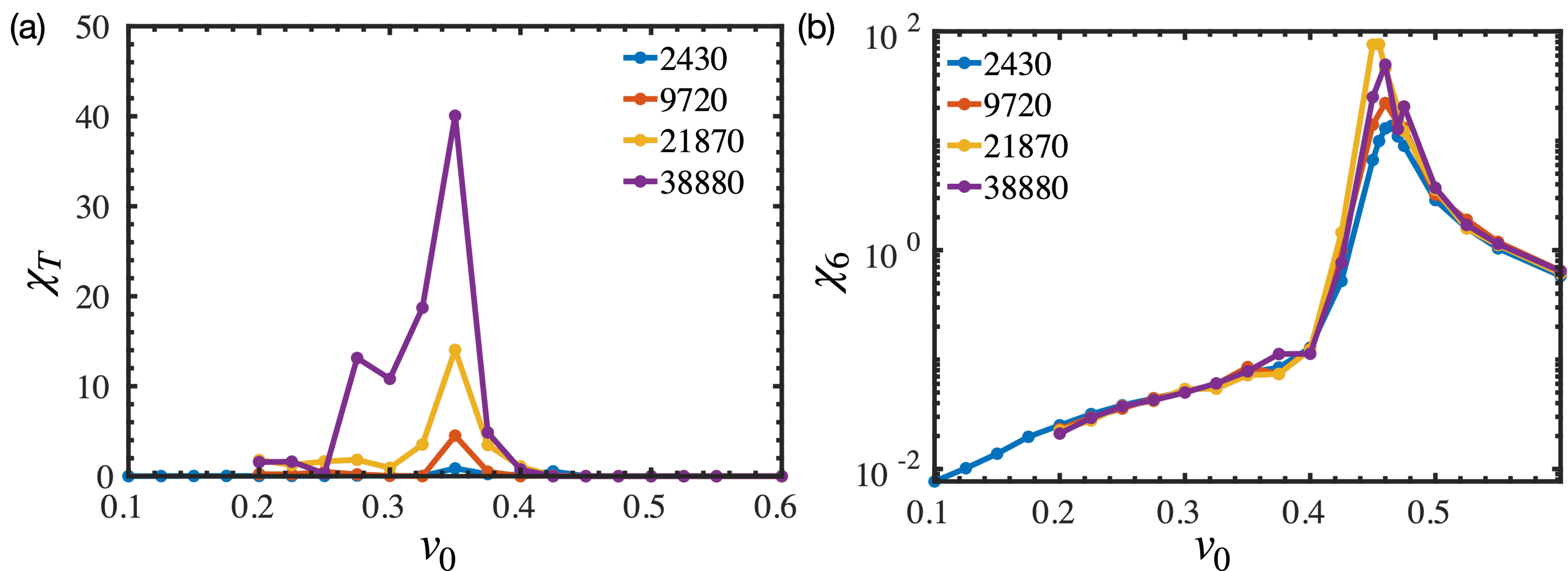}
    \caption{\textbf{The susceptibility during melting as a function of $v_0$ for various system sizes.} (a) All the $\chi_T$ in melting have a sharp peak at $v_0=0.35$, clearly indicating the crystal-hexatic transition. (b) All the $\chi_6$ have a sharp peak at $v_0=0.46\pm0.01$, clearly indicating the hexatic-liquid transition.}
    \label{fig:chi_no_div}
\end{figure}


\begin{figure}[htbp]
    \centering
    \includegraphics[width=0.5\textwidth]{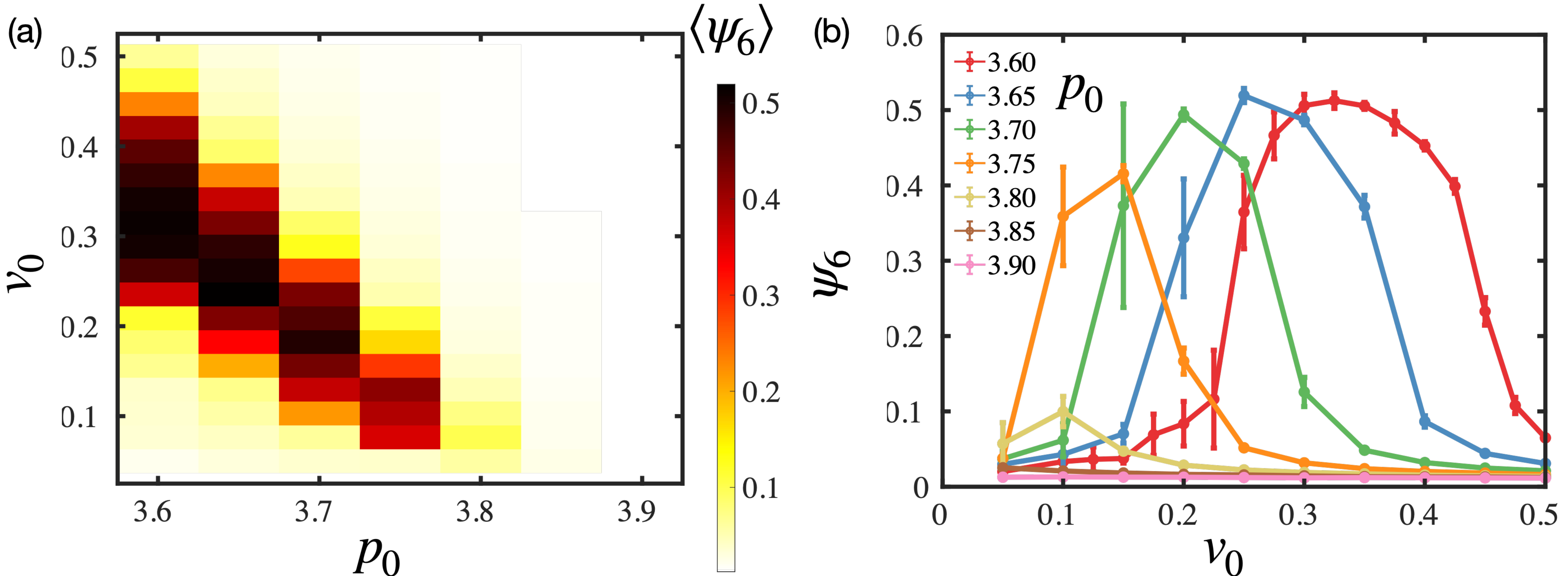}
    \caption{\yw{\textbf{Depedence of the  orientational order parameter   $\Psi_6$ on different choices of  $p_0$.}
    (a) The phase diagram as a function of cell shape index $p_0$ and motility $v_0$ with $\gamma_0=2\times10^{-5}$. The color indicates the magnitude of $\Psi_6$. 
     (b) The orientational order parameter $\Psi_6$ as the function of  $v_0$ with various $p_0$ and fixed $\gamma_0=2\times10^{-5}$. 
    The high $\Psi_6$ values at intermediate $v_0$ levels indicate the appearance of the hexatic phase. 
    Cell division leads to a liquid-hexatic-liquid transition sequence across different $p_0$ values. 
    The peak of $\Psi_6$ shrinks and shifts to smaller $v_0$ as $p_0$ increasingly approaches the $p_0\sim3.75$, which is in harmony with the no-division case\cite{pasupalakHexaticPhaseModel2020}.
    }}
    \label{fig:psi_v0_p0}
\end{figure}

\begin{figure}[htbp]
    \centering
    \includegraphics[width=0.5\textwidth]{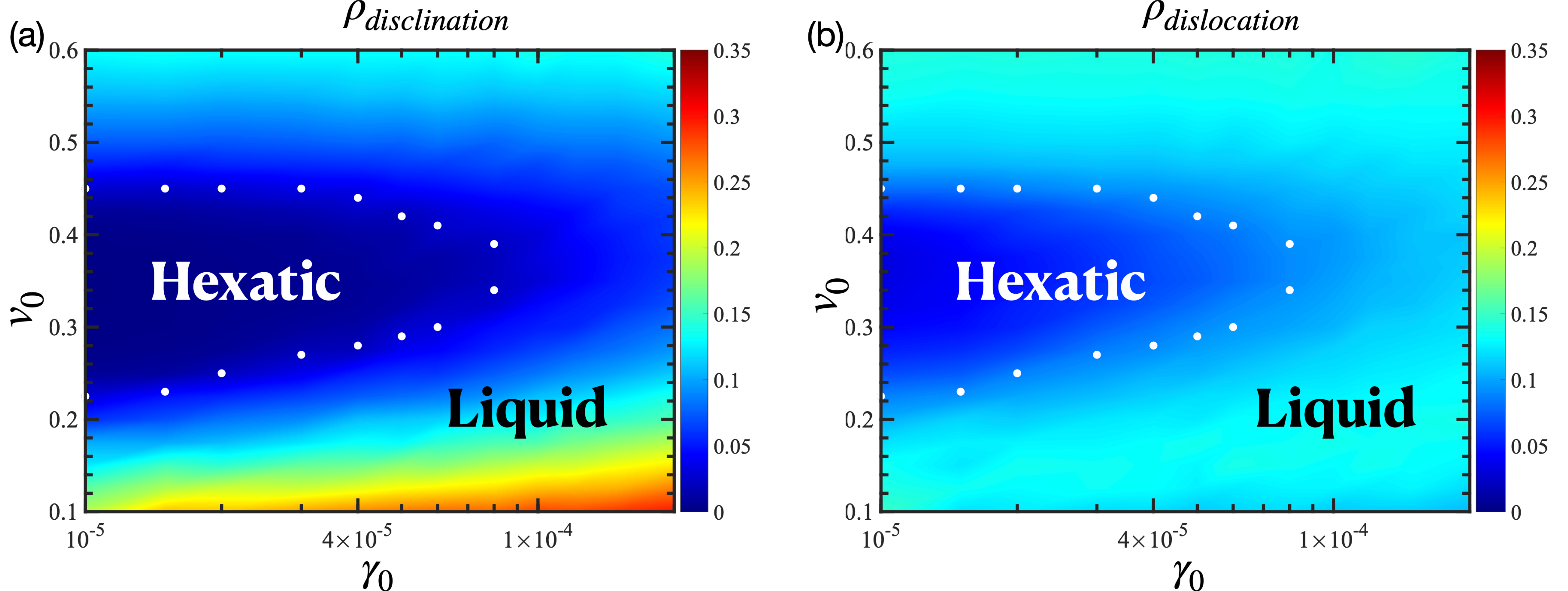}
    \caption{\textbf{The density of (a) disclinations and (b) dislocations as a function of cell division rate $\gamma_0$ and motility $v_0$.} Color indicates the magnitude of the density of (a) disclinations and (b) dislocations. White dots are the hexatic-liquid transition points, obtained from the divergence of $\chi_6$. In the hexatic phase, the density of disclinations is very low. In the liquid phase, the density of disclinations increases rapidly at low $v_0$.
    }
    \label{fig:defect_v0_gamma0}
\end{figure}

\section{Meanfield model}

{\color{black}
The mean-field approximation dynamics of a single small ($\sim 4$ cells) cell cluster can be classified into three states: 
(1) a fully ordered crystalline state (X),
(2) an isolated dislocation corresponding to a hexatic state (H),
(3) an isolated single disinclination corresponding to a liquid state (D).
The corresponding volume densities (or probabilities) are given by $\rho_X$, $\rho_H$, and $\rho_D$, respectively.

Here, the crystalline (X) state is a global energy minimum, while the hexatic (H) and liquid (D) states are local energy minima. The model assumes that one state can transition to another due to thermal fluctuations. Taken together, we model the dynamics as activated hops between energy minima (Fig.~5(a)), where the associated energy barriers are given by $\Delta\epsilon_X$, $\Delta\epsilon_H$, and $\Delta\epsilon_D$, respectively.
Since dislocations and disclinations are excitations of the crystalline state, we write $\Delta\epsilon_H = c_H\Delta\epsilon_X$ and $\Delta\epsilon_D = c_D\Delta\epsilon_X$ with $1 > c_H > c_D$. This is also well supported by previous work~\cite{bi_energy_barrier} which computed the energy barriers between four-cell clusters with different topologies. Here, rather than computing the precise values of the energy barriers, we make an arbitrary choice of $c_H=0.8$ and $c_D=0.2$ without the loss of generality.
}

 \subsection{In the absence of cell divisions, the meanfield model recapitulates the  two-step melting process} 
 In the absence of cell divisions, the cell motility provides a source of random fluctuations. In  the limit of $D_r \gg 1$, it essentially provides a source of uncorrelated fluctuations that can be described by an effective temperature given by~\cite{biMotilityDrivenGlassJamming2016}
\begin{equation}
  T \propto v_0^2,
\end{equation}
which provides the thermal-like activation to overcome barriers in the mean-field model.
The transition rates between states are determined by the thermally activated process, i.e. $R e^{-\Delta \epsilon/T}$. Here the attempt frequency $R$ between two states (assumed to be the same for all states), and the related energy barriers as illustrated in Fig.~5(a). 
The states therefore evolve according to
\begin{equation}
\label{eq:kinetics_nodiv}
\begin{aligned}
    \dot{\rho}_X  &= R e^{-\Delta\epsilon_H /T}\rho_H - R e^{-\Delta\epsilon_X/T }\rho_X,\\
    \dot{\rho}_H &= R e^{- \Delta\epsilon_X /T}\rho_X + R e^{-\Delta\epsilon_D/T}\rho_D - 2R e^{- \Delta\epsilon_H / T}\rho_H,\\
    \dot{\rho}_D  &= R e^{-\Delta\epsilon_H/T}\rho_H - R e^{-\Delta\epsilon_D/T}\rho_D.
\end{aligned}
\end{equation}

\begin{figure}[htbp]
    \centering
    \includegraphics[width=0.5\textwidth]{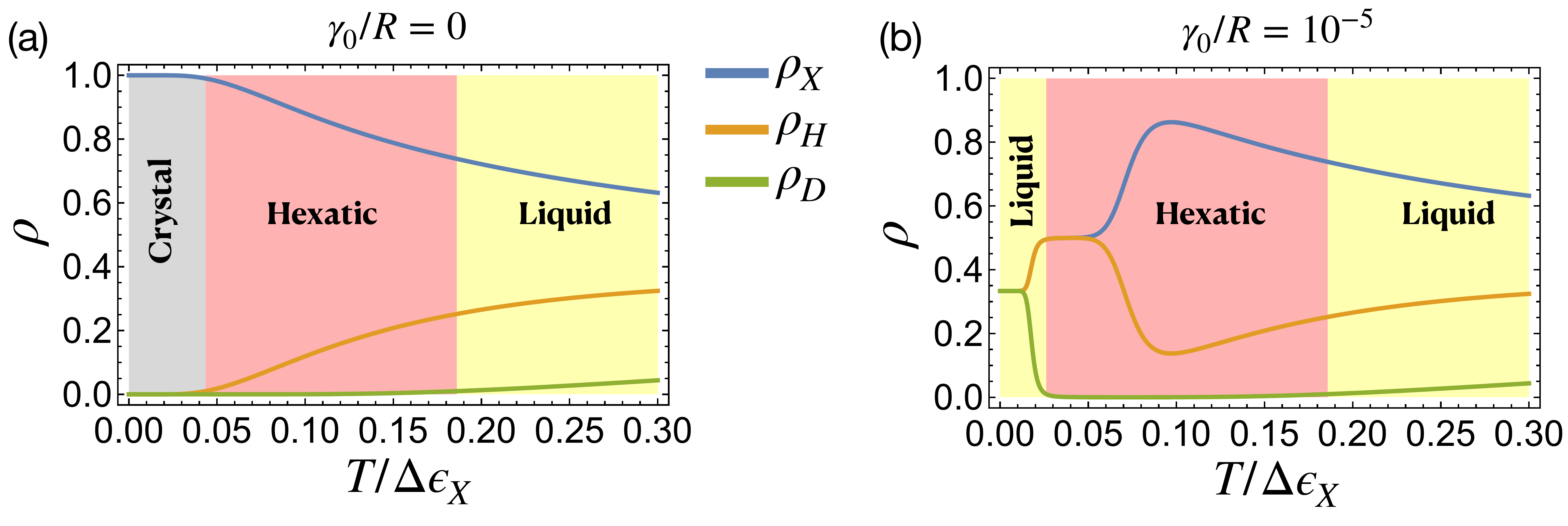}
    \caption{
    \yw{(a) Plot of $\rho_X$, $\rho_H$ and $\rho_D$ as a function of temperature in the absence of cell division as obtained by the meanfield model. Here, the tissue undergoes a two-step melting process, transitioning from a crystalline structure to a liquid state through an intermediate hexatic phase.
    (b) In the presence of cell divisions ($\gamma_0/R =10^{-5}$), the crystalline state ceases to exist across the entire temperature range, giving rise to a re-entrant  Liquid - Hexatic - Liquid transition.}
     }
    \label{fig:MF_rho_t}
\end{figure}

\yw{
The steady-state solution gives $\rho_X = Q^{-1} e^{\Delta\epsilon_X/T}$, $\rho_H = Q^{-1} e^{\Delta\epsilon_H/T}$, and $\rho_D = Q^{-1} e^{\Delta\epsilon_D/T}$, where the normalization factor is given by $Q = e^{\Delta\epsilon_X/T} + e^{\Delta\epsilon_H/T} + e^{\Delta\epsilon_D/T}$. Choosing $1/R$ as the unit of time and $\Delta\epsilon_X$ as the unit of energy, we plot the fraction of the three states as a function of the effective temperature in Fig.~\ref{fig:MF_rho_t}(a). 
As the temperature increases, the model exhibits a two-step melting process. The system first transitions from a crystalline phase to a hexatic phase and then from hexatic to liquid. We can define each phase by selecting a threshold value as follows:
(1) Crystalline: For values where $\rho_X > 0.01$ and both $\rho_H$ and $\rho_D$ are less than $0.01$.
(2) Hexatic: When $\rho_H > 0.01$ and $\rho_D < 0.01$.
(3) Liquid: In cases where $\rho_D > 0.01$.
It is important to note that the results are independent of the precise threshold value chosen.
}

\subsection{Cell division/apoptosis alters the melting process }
Cell division and apoptosis introduce a new type of active force. The resulting active cell-shape deformations enhance the fluctuations over energy barriers by `tilting' the energy landscape. The transition rates are modified by adding a $\gamma_0$ term as shown in Fig.~5(a).
The master equations then become
\begin{equation}
\label{eq:kinetics_div}
\begin{aligned}
    \dot{\rho}_X   =& (Re^{-\Delta\epsilon_H/T}+\gamma_0)\rho_H-(Re^{-\Delta\epsilon_X/T}+\gamma_0)\rho_X\\
    \dot{\rho}_H   =& (Re^{-\Delta\epsilon_X/T}+\gamma_0)\rho_X+(Re^{-\Delta\epsilon_D/T}+\gamma_0)\rho_D
     -2(Re^{-\Delta\epsilon_H/T}+\gamma_0)\rho_H\\
    \dot{\rho}_D   =& (Re^{-\Delta\epsilon_H/T}+\gamma_0)\rho_H-(Re^{-\Delta\epsilon_D/T}+\gamma_0)\rho_D
\end{aligned}
\end{equation}
{\color{black}
The steady-state solution gives $\rho_X=Q^{-1}(e^{-\Delta\epsilon_X/T}+\gamma_0)^{-1}$, 
$\rho_H=Q^{-1}(e^{-\Delta\epsilon_H/T}+\gamma_0)^{-1}$ 
and $\rho_D=Q^{-1}(e^{-\Delta\epsilon_D/T}+\gamma_0)^{-1}$ with $Q=(e^{-\Delta\epsilon_X/T}+\gamma_0)^{-1}+(e^{-\Delta\epsilon_H/T}+\gamma_0)^{-1}+(e^{-\Delta\epsilon_D/T}+\gamma_0)^{-1}$. 
Fig.~\ref{fig:MF_rho_t}(b) shows $\rho_X$, $\rho_H$ and $\rho_D$ as a function of temperature at a constant division rate ($\gamma_0/R =10^{-5}$). In the limit of high temperature, the model behaves similarly to the case of no division. However, the dominance of the division term in the transition rates dramatically changes the steady-state solution in the low-temperature limit, driving the tissue from a crystal phase to a liquid phase.

In  Figure 5(b) of the main text, we present the phase diagram, plotted in terms of the dimensionless variables $\gamma_0/R$ and $\sqrt{T/\Delta\epsilon_X}$. This diagram is constructed using a uniform threshold criterion across the three distinct phases. This illustrates that mean-field theory effectively replicates the behavior evident in our numerical simulations, specifically within the experimentally accessible range of cell division rates($ \gamma_0/R > 10^{-5}$). 
}

 \begin{figure}[t]
    \centering
    \includegraphics[width=0.5\textwidth]{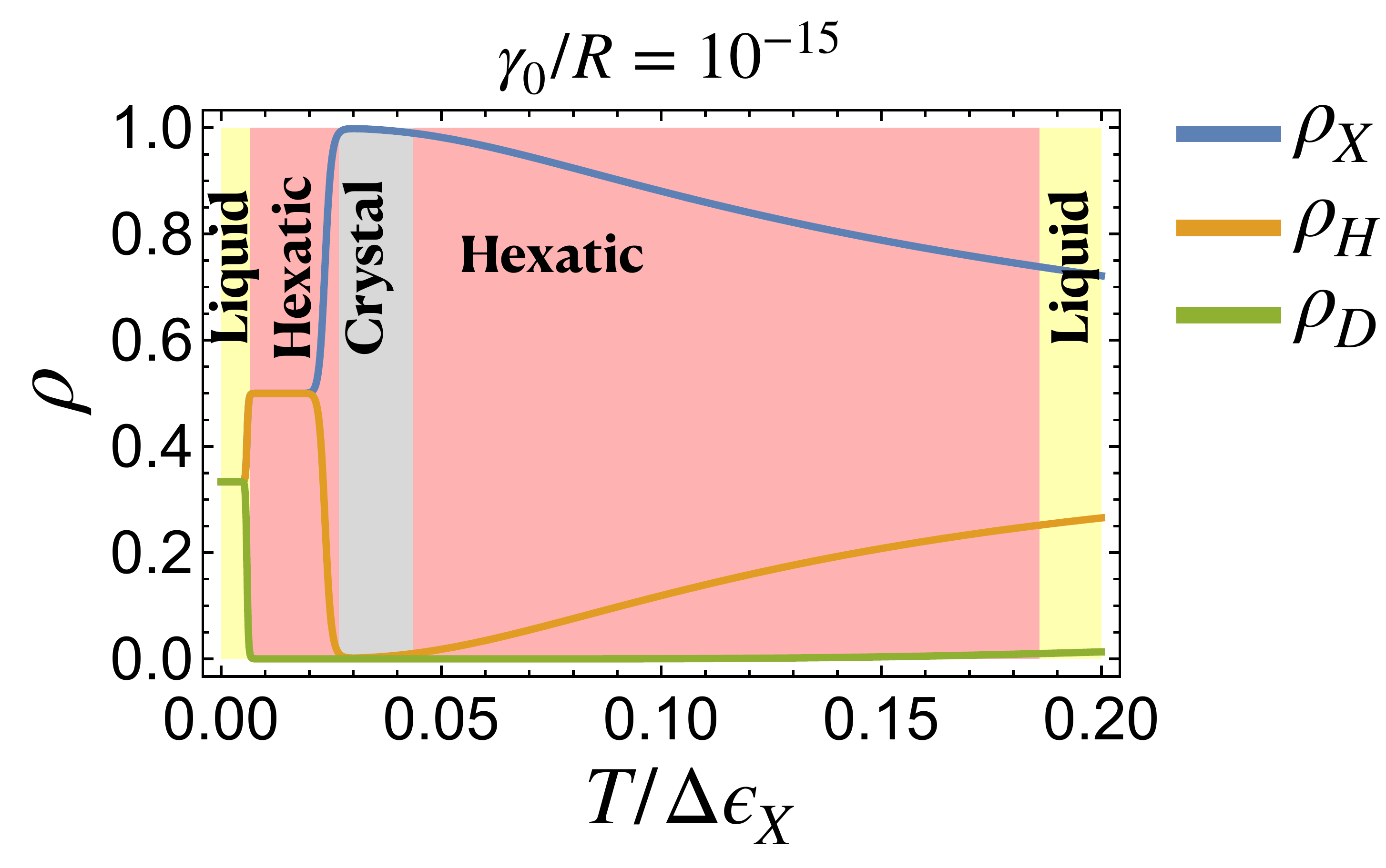}
    \caption{
    The behavior of the three state densities as functions of temperature for a constant, ultra-low choice of division rate $\gamma_0/R = 10^{-15}$. 
    }
    \label{fig:mf_ultra_low}
\end{figure}
 \begin{figure}[htbp]
    \centering
    \includegraphics[width=0.45\textwidth]{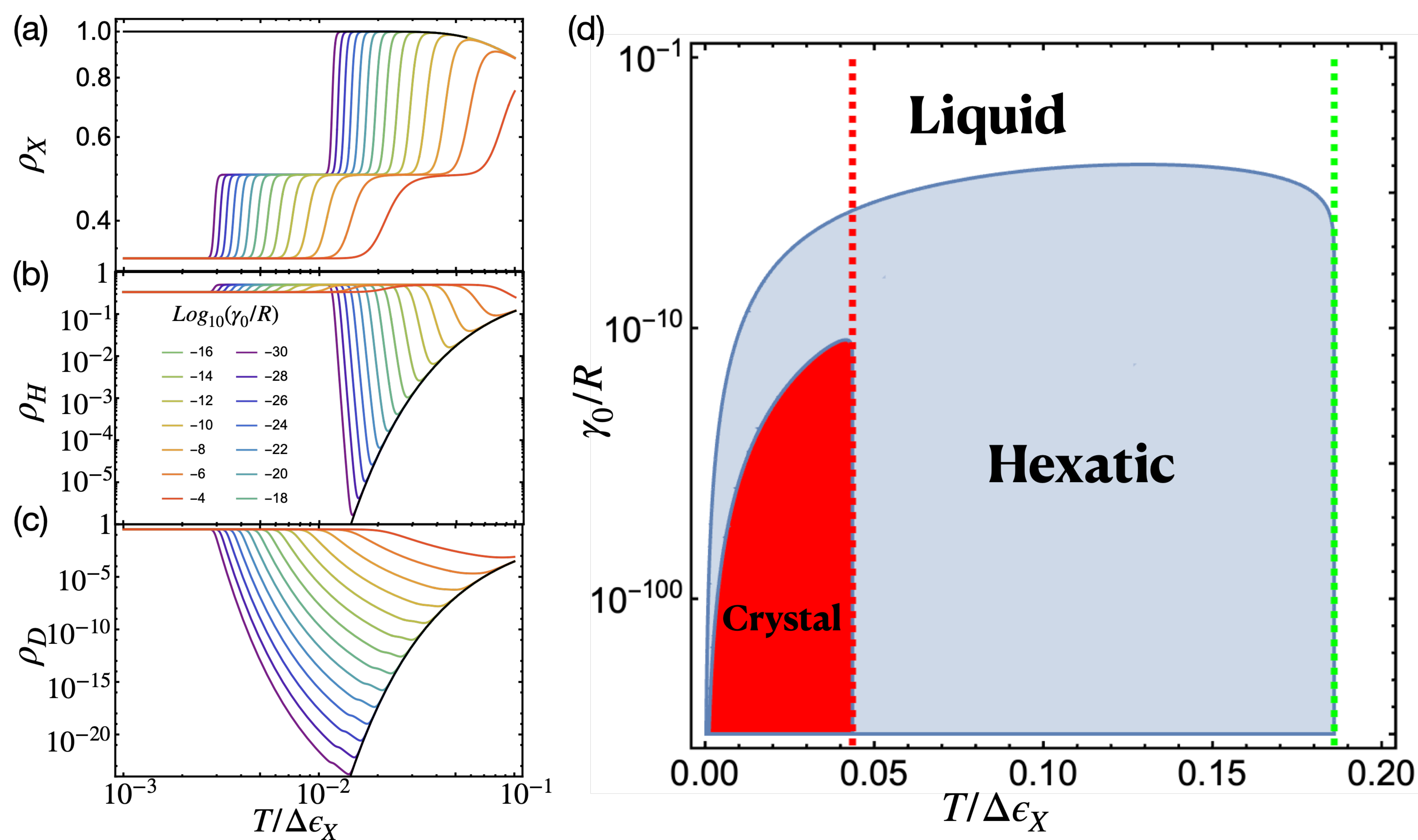}
    \caption{
    The state densities (a) $\rho_X$ (b) $\rho_H$ (c) $\rho_D$ as the function of dimensionless temperature $T/\Delta\epsilon_X$ under the $\gamma_0\rightarrow0$ limit. As the division decreases to ultra-low values, the state densities share the same steady solution with the no-division case in a bigger range with a lower limit of $T/\Delta\epsilon_X$. However, the tissue is inevitably in a liquid phase at the zero temperature limit, having equal $\rho_X$, $\rho_H$, and $\rho_D$. The possible crystal or hexatic phase can only happen in the middle-temperature range.
    The dimensionless division rate $10^{-4}, 10^{-6},...10^{-30}$ are represented by the rainbow color from red to purple. 
    (d) A re-plot of the phase diagram of the MF model where the cell division rates are displayed on a ``log of log" scale to emphasize the behavior at extremely slow values. 
    }
    \label{fig:MF_gamma20}
\end{figure}

\subsection{Behavior at ultra-low division rates}

Intriguingly, the theory also makes a prediction for ultra-low division rates ($ \gamma_0/R \ll 10^{-10}$). Here, tissues would experience a notably complex transition sequence as a function of temperature, following a Liquid-Hexatic-Crystal-Hexatic-Liquid path at constant $\gamma_0/R$ as shown in Fig. 5(b). We also show one such transition path for $\gamma_0/R = 10^{-15}$ in Fig.~\ref{fig:mf_ultra_low}. 

In order to delve deeper into the behavior as  $ \gamma_0/R \to 0$, we examine the state densities as functions of temperature for $ \gamma_0/R$  values smaller than $ 10^{-10} $, as illustrated in Fig.~\ref{fig:MF_gamma20}(a-c).
Firstly,  in the ultra-low regime of $ \gamma_0/R $, there exists a condition for a crystalline phase at sufficiently low temperatures. Under this condition, the transition rate from the Crystal phase to the Hexatic phase exceeds that from the Hexatic phase to the Crystal phase. Simultaneously, the transition rate from the Dislocation phase to the Hexatic phase is also higher than the rate from the Hexatic phase to the Dislocation phase. This condition is given by
$$
\frac{e^{-\Delta\epsilon_H/T}+\gamma_0/R}{e^{-\Delta\epsilon_X/T}+\gamma_0/R}
 < 1 \quad \text{and} \quad
 \frac{e^{-\Delta\epsilon_H/T}+\gamma_0/R}{e^{-\Delta\epsilon_X/T}+\gamma_0/R}
 < 1.
$$
As the ratio $\gamma_0/R$ decreases further, the stable crystalline phase not only emerges at specific temperatures (as shown in Fig.~\ref{fig:mf_ultra_low}), but also expands its temperature range until it ultimately converges with the crystalline phase present at $\gamma_0/R=0 $.

Finally, we summarize these behaviors in the phase diagram (Fig.~\ref{fig:MF_gamma20}(d)), which is the same as Fig.~5(b), except the $\gamma_0/R$ here is plotted on a "log-of-log" scale to emphasizes the slow convergence of the quantity. 
The phase diagram of the MF model exhibits a striking resemblance to that of the 2D melting on a substrate with quenched random disorder in its topography, as discussed in work by Sachdev and Nelson ~\cite{sachdev1984}. In both cases, temperature acts as the driving force for the phase transition. Interestingly, in our model, cell division serves a role analogous to the degree of quenched disorder in~\cite{sachdev1984}. This correlation is logical since cell division introduces spatially random distortions in the tissue, akin to the disorder in the substrate, and these effects persist without annealing over time, forming a quenched random variable.

\end{document}